\documentclass[fleqn,usenatbib,useAMS]{mnras}
\usepackage{amssymb,amsmath}
\usepackage{graphicx,epstopdf}
\usepackage{soul,xcolor}

\newcommand{\m}[1]{\boldsymbol{#1}}

\title[The force-free twisted magnetosphere -- II]{The force-free twisted magnetosphere of a neutron star\,--\,II: \\ Degeneracies of the Grad--Shafranov equation}
\author[T. Akg\"{u}n et al.]
{T.~Akg\"{u}n$^1$\thanks{E-mail: akgun@astro.cornell.edu}, P.~Cerd\'{a}--Dur\'{a}n$^2$, J.A.~Miralles$^1$, and J.A.~Pons$^1$ 
\\$^1$Departament de F\'{i}sica Aplicada, Universitat d'Alacant, Ap. Correus 99, 03080 Alacant, Spain
\\$^2$Departament d'Astronomia i Astrof\'{i}sica, Universitat de Val\`{e}ncia, Dr. Moliner 50, 46100, Burjassot, Val\`{e}ncia, Spain}

\begin{document}
\label{firstpage}
\pagerange{\pageref{firstpage}--\pageref{lastpage}}
\maketitle

\begin{abstract}
We extend our previous study of equilibrium solutions of non-rotating force-free magnetospheres of neutron stars. We show that multiple solutions exist for the same sets of parameters, implying that the solutions are degenerate. We are able to obtain configurations with disconnected field lines, however, in nearly all cases these correspond to degenerate higher energy solutions. We carry out a wide parametric search in order to understand the properties of the solutions. We confirm our previous results that the lower energy solutions have up to $\sim 25\%$ more energy than the vacuum case, helicity of the order of $\sim 5$ (in some defined units), maximum twist of $\sim 1.5$\,rad, and a dipole strength that is up to $\sim 40\%$ larger than the vacuum dipole. Including the degenerate higher energy solutions allows for larger theoretical limits of up to $\sim 80\%$ more energy with respect to the vacuum case, helicity of the order of $\sim 8$, and a dipole strength that can now be up to four times that of the vacuum dipole, while the twist can be significantly larger and even diverge for configurations with disconnected domains. The higher energy solutions are probably unstable, therefore, it is unlikely that such magnetospheres exist under normal conditions in magnetars and high magnetic field pulsars.
\end{abstract}



\section{Introduction}
Magnetars are strongly magnetized neutron stars with inferred magnetic fields above $10^{14}$\,G. One of the most distinctive features of these objects is the recurrent X-ray activity in the form of short duration bursts ($10^{36}-10^{43}$\,erg/s in $\sim 0.1$\,s), long duration outbursts ($10^{36}$\,erg/s in hours), and energetic giant flares ($10^{44}-10^{47}$\,erg/s in $\sim 0.1$\,s) \citep{2015SSRv..191..315M, 2017ARA&A..55..261K}. This activity may be accompanied by an extended X-ray decay lasting $10^3-10^4$ times longer than the original event \citep{2017MNRAS.471.1819C}. The exact origin of this activity is currently unknown but is clearly linked to the presence of a strong magnetic field, which is slowly evolving mainly due to the dominant effect of the Hall drift and Ohmic dissipation in the crust \citep{Jones1988,1992ApJ...395..250G,Pons:2009,Gourgouliatos:2016}. Two competing (or more precisely, complementary) models currently try to explain the triggering of these violent events. In the classical model, Hall drift of magnetic field lines builds up stresses in the stellar crust that eventually lead to a mechanical failure \citep{Thompson:1996,Perna:2011}. This sudden ``crustquake'' releases energy and immediately disturbs the magnetosphere, producing the observed X-ray emission. On the other hand, in more recent works, it is assumed that the crust does not break suddenly, but yields elastically to Hall-induced stresses up to a certain point, beyond which it deforms plastically \citep{2014ApJ...794L..24B,2017ApJ...841...54T}. As that happens and the footprints of the magnetic field lines at the surface are slowly displaced, there is a transfer of helicity into the magnetosphere, leading to a magnetic reconnection event when some maximum twist is reached \citep{Lyutikov:2003,Gill:2010,2013ApJ...774...92P,2017MNRAS.472.3914A}. Regardless of the details of the triggering model, it is clear that the X-ray emission is a consequence of the presence of a strongly twisted magnetosphere, which is potentially prone to severe magnetic instabilities. Therefore, it is important to study the equilibrium of magnetospheres. As rotation has negligible effects for typical magnetar spin periods, it can be neglected.

Equilibrium solutions of force-free twisted magnetospheres have been considered in the past by a number of authors \citep{2014MNRAS.445.2777F,2014MNRAS.437....2G,2015MNRAS.447.2821P,2016MNRAS.462.1894A,2017MNRAS.468.2011K}, solving the Grad--Shafranov (GS) equation matched to the magnetic field at the neutron star surface. If one starts with an untwisted current-free magnetosphere (i.e.\ a potential solution) and computes a sequence of force-free (but no longer current-free) equilibria with increasing twist, the resulting magnetosphere tends to inflate as the energy and helicity stored in the magnetosphere increase \citep[see e.g.][]{1995ApJ...443..810W}. For sufficiently large twists, \cite{2014MNRAS.445.2777F}, \cite{2014MNRAS.437....2G}, \cite{2015MNRAS.447.2821P} and \cite{2017MNRAS.468.2011K} observed the formation of disconnected loops of current in the magnetosphere. However, these kind of configurations were not found in \cite{2016MNRAS.462.1894A} (\emph{Paper I}, hereafter) for a comparable set of parameters. Instead, beyond a maximum twist of $\sim 1.5$\,rad, the numerical procedure used to solve the non-linear GS equation failed to converge. This maximum twist was interpreted as a physical bound of the system and not as a limitation of the numerical solver (see Paper I), and can be used to estimate the twist at which reconnection events are produced during the magneto-thermal evolution of magnetars \citep{2017MNRAS.472.3914A}. Furthermore, the values of the maximum twist obtained with this procedure coincide with those obtained in dynamical MHD simulations \citep{Mikic:1994,2012ApJ...754L..12P,2013ApJ...774...92P}.

In this work, we study in more detail the force-free configurations of Paper I in order to try to understand the differences with other authors in the appearance of disconnected regions in the magnetosphere. Our hypothesis, which is verified in this work, is that the solutions of the GS equation are degenerate, and that beyond a certain critical twist, the solution of this equation is non-unique. For this purpose we use an iterative numerical procedure similar to that of \cite{2015MNRAS.447.2821P}. We show that, for a given set of parameters resulting in multiple solutions, the solutions presented in Paper I correspond to the lowest energy (and helicity and twist) and hence should be considered as the stable branch of magnetospheric configurations.

The structure of the paper is as follows: in \S\ref{section_overview} we present a short statement of the problem, reviewing the most relevant equations, and then we discuss the new method applied for solving the equation iteratively; in \S\ref{section_results} we discuss our findings and their implications; and in \S\ref{section_conclusions} we present our conclusions.

\section{The Grad--Shafranov equation}\label{section_overview}
The nature of the problem and the numerical methods we employ are discussed in great length in Paper I. Here, we only give a minimal review of the related equations as a reference.

We express the axisymmetric magnetic field in terms of the poloidal stream function $P$ and the toroidal stream function $T$ as,
	\begin{equation}
	\m{B} = \m{\nabla} P \times \m{\nabla} \phi + T \m{\nabla} \phi \ ,
	\end{equation}
$\phi$ being the azimuthal angle. The vanishing of the azimuthal component of the Lorentz force implies that $T$ must be a function of $P$. Setting the remaining poloidal component of the force equal to zero then gives the GS equation\footnote{Force-free fields have been studied in \cite{1954ZA.....34..263L}, prior to the works by Grad and Shafranov.}
	\begin{equation}
	\triangle_{\rm GS} P = - G(P)  \ ,
	\label{eq:GS}
	\end{equation}
where $G(P) = T(P) T'(P)$ and $\triangle_{\rm GS}$ is the GS operator given by
	\begin{equation}
	\triangle_{\rm GS} = \varpi^2 \m{\nabla} \cdot (\varpi^{-2} \m{\nabla})
	= \partial_r^2 + \frac{1 - \mu^2}{r^2} \partial_\mu^2 \ ,
	\end{equation}
with the notation $\varpi = r\sin\theta$ and $\mu=\cos\theta$.

We assume a toroidal function of the form,
	\begin{equation}
	T(P) = \left\{
	\begin{aligned}
	& s(P-P_{\rm c})^\sigma & \mbox{for} \ P \geqslant P_{\rm c} \ , \\
	& 0 & \mbox{for} \ P < P_{\rm c} \ .
	\end{aligned}
	\right.
	\label{toroidal}
	\end{equation}
Here, $s$ controls the relative strength of the toroidal field with respect to the poloidal field; $P_{\rm c}$ is the critical field line defining the border of the toroidal field (i.e.\ the region of magnetospheric currents); and $\sigma$ controls how the toroidal field amplitude is concentrated within this region. Regularity of currents requires $\sigma \geqslant 1$. Continuity further requires $\sigma > 1$, and as noted in Paper I, the case of $\sigma = 1$ implies that the current has a sudden jump (discontinuity) at the border of the toroidal field, while the magnetic field, obviously, is continuous (as the poloidal field is tangential to this border, and the toroidal field goes to zero). This discontinuity does not correspond to surface currents, which would be problematic as they would result in discontinuities in the magnetic field. Increasing $\sigma$ lessens the effect of the toroidal field. Therefore, in order to work with the more extreme case, we choose $\sigma = 1$, as we did in Paper I. In reality, taking a slightly larger value, for example $\sigma = 1.1$ as in \cite{2009MNRAS.395.2162L}, makes little difference, as shown in Paper I.

\subsection{Non-uniqueness of the solutions}
The GS equation is a second-order, non-linear, inhomogeneous partial differential equation (PDE). The non-linearity of the equation is due to the presence of the function $G(P)$ on the right-hand side of equation (\ref{eq:GS}). Even in the case of $\sigma=1$, the non-linearity arises due to the discontinuous behaviour of $G(P)$ and its derivatives at $P=P_{\rm c}$. Given the elliptic nature of the equation, it is necessary to impose boundary conditions in order to obtain a solution. In our case, boundary conditions are completely set for fixed values of the parameters $\sigma$, $s$ and $P_{\rm c}$, which effectively determine the matching condition at the stellar surface as a Dirichlet boundary condition (see Paper I for details). However, even if the freedom at the boundary is completely set, the existence and uniqueness of the solutions of the equation cannot be taken for granted.

The customary way of demonstrating uniqueness of the solutions of elliptic equations is to find a maximum principle. Equation (\ref{eq:GS}) is quasi-linear, because it is linear in its second (highest) derivatives. For such equations of the form
	\begin{equation}
	\Delta u = f(u),
	\end{equation}
$\Delta$ being the Laplace operator and $f$ a non-linear function, it is possible to use a maximum principle to prove local uniqueness of the solution if $f'(u)\ge 0$ \citep[see][chapter 14]{taylor10}. The GS equation can be rewritten in such a form through
	\begin{equation}
	\Delta \left ( \frac{P \sin{\phi}}{r \sin \theta} \right )
	= - \frac{G (P) \sin{\phi}}{r \sin{\theta}},
	\end{equation}
where, we can now identify $u = P \sin{\phi}/(r \sin{\theta})$ and $f = - G(P) \sin{\phi} / (r \sin{\theta})$ \citep[see][]{2014MNRAS.437....2G}. In the linear case, $f \propto u$, implying that $G \propto T \propto P$, and, therefore, $f'(u) = - G'(P)$ is a non-positive constant. Even under this simplistic assumption, the only case in which we can guarantee the uniqueness of the solution of the GS equation is the current-free case ($G=0$).


More generally, for sufficiently small values of $T'(P)$, \cite{bineau1972} proved the uniqueness of force-free solutions, provided the solution domain is bounded and the field is not vanishing anywhere. However, little more is known from an analytical point of view on the conditions necessary to obtain unique solutions of the GS equation, or, in general, of three-dimensional (3D) force-free configurations. \citep[See, e.g.][for a discussion in the context of solar force-free fields.]{Wiegelmann2012}

Given that for small or no currents we expect a unique solution and for larger twists this may not be the case, we expect that there will be a critical twist beyond which no unique solution can be found. As we show in this paper, this indeed is the case.

\subsection{New numerical implementation}
We cannot use the same implementation as in Paper I in order to study the multiple solutions. The numerical procedure in that work consists of solving the linear part of the GS equation (the left-hand side of equation \ref{eq:GS}) for an initial guess for the right-hand side. This procedure is repeated while maintaining the values of $s$ and $P_{\rm c}$ fixed throughout the iterations. With this procedure, if multiple solutions exist for the same values of $s$ and $P_{\rm c}$, the numerical procedure converges to one of the solutions. The solution found depends on the initial guess, so this numerical procedure presents difficulties to obtain systematically all the solutions of the GS equation for given $s$ and $P_{\rm c}$.

In the present work, we follow the scheme devised by \cite{2015MNRAS.447.2821P}. Instead of fixing $P_{\rm c}$ during the iterations, we fix the value of the \emph{critical radius} $r_{\rm c}$, defined as the radial extent (on the equatorial plane) of the magnetic field line with stream function $P_{\rm c}$. This means that $P_{\rm c}$ varies between iterations, while the border of the toroidal region is forced to pass through $r_{\rm c}$, which is constant. As we show in the next section, for a given value of $P_{\rm c}$ there may exist multiple solutions with different values of $r_{\rm c}$. Therefore, this procedure allows us to build selectively the different configurations corresponding to a single value of $P_{\rm c}$.

We build sequences of models starting with the untwisted (current-free) solution and increasing the twist progressively by increasing the value of $r_{\rm c}$ for a fixed value of $s$. To speed up the convergence of our iterative scheme, we use the solution of the previous model as an initial guess for the subsequent model in the sequence.

\subsection{Notation and units}
In this work, we use the same notation and dimensionless units as in Paper I. Thus, distances are measured in units of the stellar radius $R_\star$, and magnetic field strength is measured in units of some $B_{\rm o}$. For a dipole field,  $B_{\rm o}$ corresponds to the surface magnetic field strength at the equator (or equivalently, half of the magnetic field strength at the pole). The dimensions of all other quantities used in the paper can be derived from these two definitions. Thus, for example, the poloidal function $P$ is given in units of $B_{\rm o} R_\star^2$. For a dipole field, we have $P(r,\theta) = r^{-1} \sin^2\theta$, implying that, on the stellar surface (at $r = 1$), $P$ ranges from 0 (at the pole) to 1 (at the equator). The most important quantities and their units are listed in Table \ref{table_dimensions} as a reference.

	\begin{table}
	\caption{List of relevant quantities, notation and units.}
	\label{table_dimensions}
	\center
	\begin{tabular}{lcl}
		\hline \hline
		Quantity & Notation & Units \\
		\hline \hline
		Magnetic field strength & $B$ & $B_{\rm o}$ \\
		Radius & $r$ & $R_\star$ \vspace{0.15cm} \\
		Poloidal function & $P$ & $B_{\rm o} R_\star^2$ \\
		Toroidal function & $T$ & $B_{\rm o} R_\star$ \vspace{0.15cm} \\
		Energy & ${E}$ & $B_{\rm o}^2 R_\star^3$ \\
		Helicity & ${H}$ & $B_{\rm o}^2 R_\star^4$ \\
		Twist & $\varphi$ & ${\rm rad}$ \\
		\hline \hline
	\end{tabular}
	\end{table}


\section{Results}\label{section_results}
In Paper I, we conjectured that the disconnected field lines reported by \cite{2015MNRAS.447.2821P} might represent degenerate solutions of the GS equation. Here, we employ the term \emph{degenerate} in the sense that multiple magnetospheric solutions can be constructed for the same values of the parameters $s$ and $P_{\rm c}$ (but with different $r_{\rm c}$). Using a similar iteration scheme to \cite{2015MNRAS.447.2821P}, we now present results that effectively confirm this hypothesis. Since we fix $r_{\rm c}$ and let $P_{\rm c}$ to vary, the best way to carry out a detailed parametric study is to follow the progression of $P_{\rm c}$ for a given value of $s$, as $r_{\rm c}$ is gradually moved away from the stellar surface (which corresponds to $r_{\rm c} = 1$ in the dimensionless units listed in Table~\ref{table_dimensions}). We indeed find that there are multiple solutions for the same set of $s$ and $P_{\rm c}$ for different values of $r_{\rm c}$. In fact, in some cases we find up to three solutions, and it is likely that the progression can be continued further. In practice, for larger values of $r_{\rm c}$, convergence becomes increasingly tedious, requiring exceedingly good initial guesses, and greatly enhancing the resolution is impractical.

We next present some sample models and then explore the parameter space in greater detail. In all cases considered in this paper, we impose a dipole field at the stellar surface, $P_{\rm surf} \equiv P(1,\theta) = \sin^2\theta$ (where $r = 1$ corresponds to the stellar radius in the units employed here).

\subsection{Sample field configurations}
\begin{figure}
		\centerline{\includegraphics[width=0.5\textwidth]{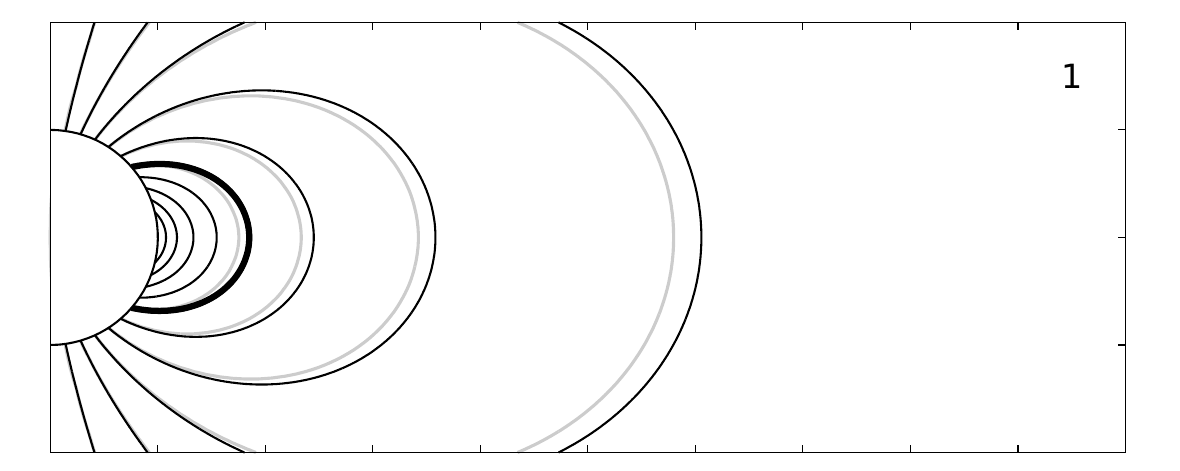}}
		\centerline{\includegraphics[width=0.5\textwidth]{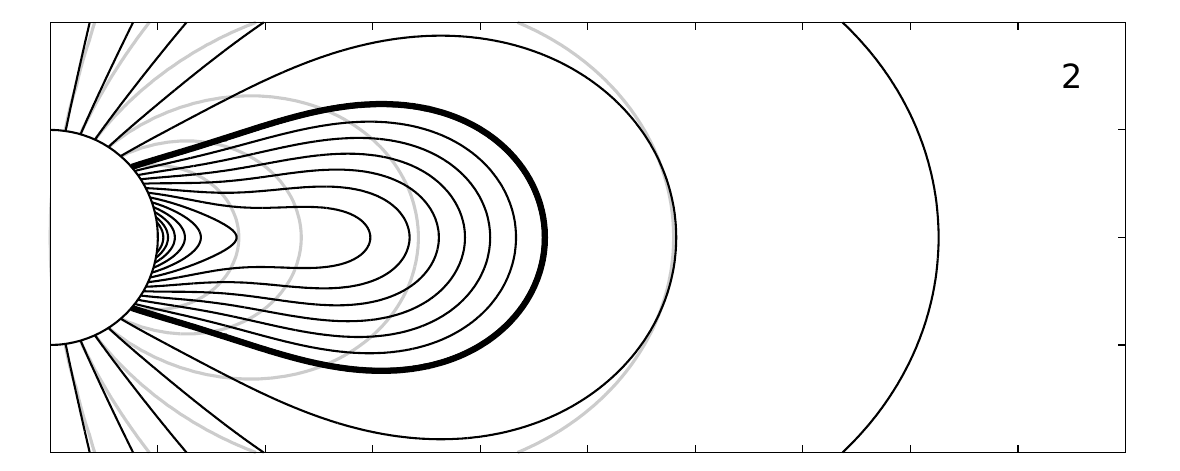}}
		\centerline{\includegraphics[width=0.5\textwidth]{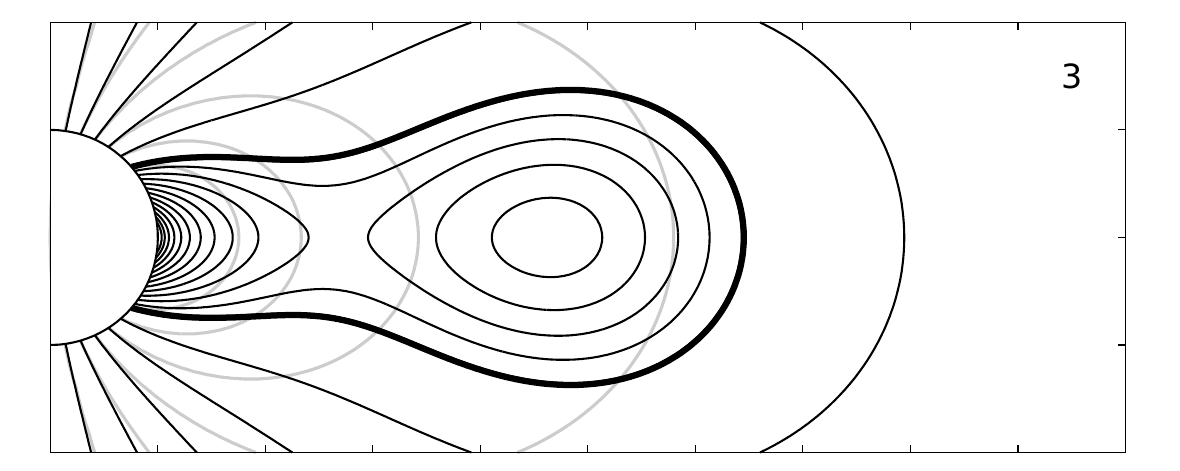}}
		\caption{Sample field configurations in 2D for $s = 1.5$ and $P_{\rm c} \approx 0.57$. Top: $r_{\rm c} = 1.85$ (lowest energy solution, also found by fixing $P_{\rm c}$). Middle: $r_{\rm c} = 4.60$ (highest energy solution). Bottom: $r_{\rm c} = 6.45$ (intermediate energy solution). The toroidal field is confined within the critical field line indicated by a thick black line. The vacuum dipole field is shown in the background in gray lines for reference.}
		\label{fig_contour}
	\end{figure}

\begin{figure}
		\centerline{\includegraphics[width=0.5\textwidth]{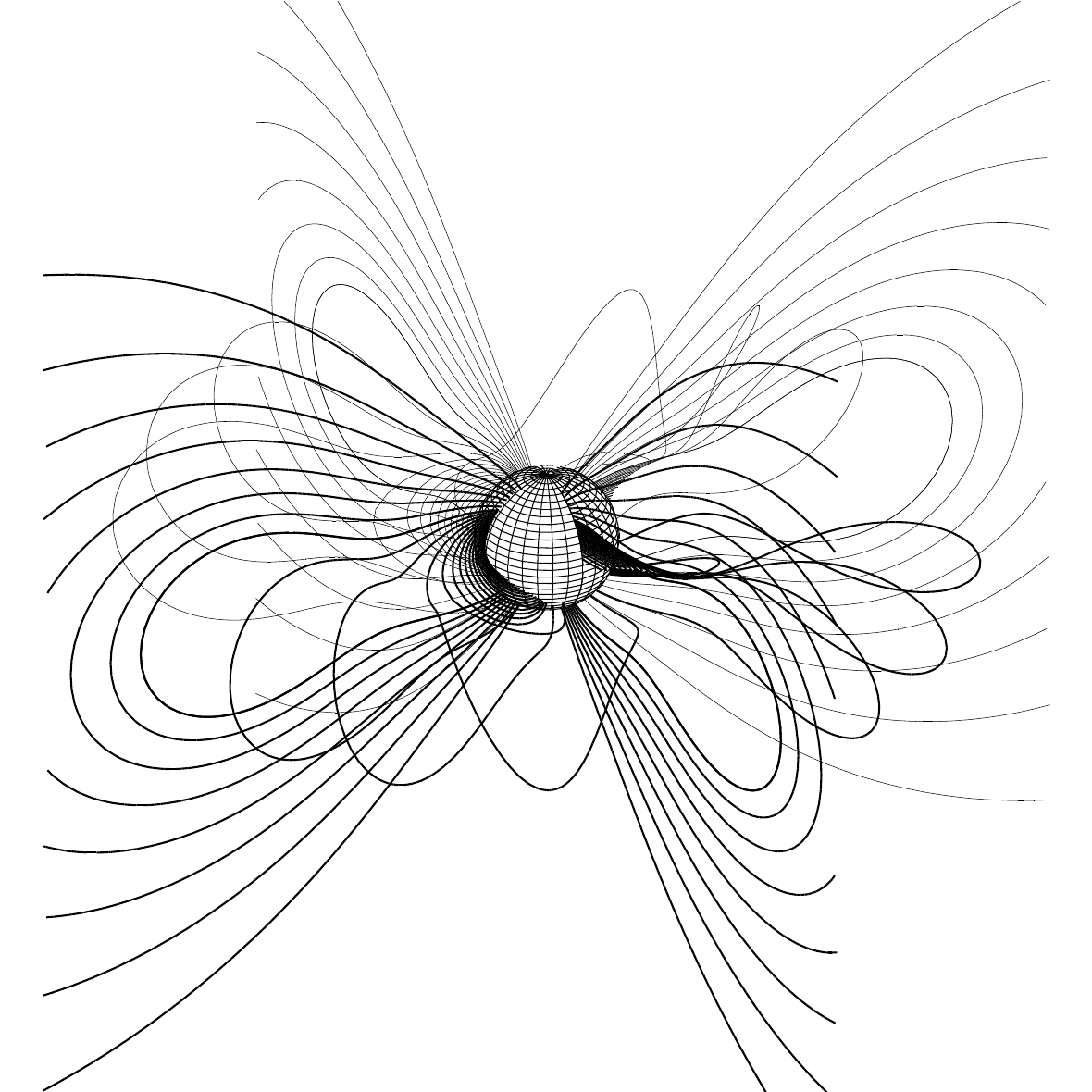}}
		\caption{3D field configuration for a sample field. (Model 3 shown in the bottom panel in Fig.~\ref{fig_contour}.)}
		\label{fig_3d}
	\end{figure}

In Fig.~\ref{fig_contour}, we show the 2D field configurations of three degenerate solutions for $s = 1.5$ and $P_{\rm c} \approx 0.57$. The three solutions (labeled as models 1, 2 and 3) correspond to $r_{\rm c} = 1.85$, $4.60$ and $6.45$, respectively. The potential (current-free) solution for the same surface field (i.e.\ the vacuum dipole) is shown in gray lines in the background as a reference. These solutions are three representative cases of typical geometries that can be obtained for the same parameters $s$ and $P_{\rm c}$: (i) a nearly potential solution, where magnetospheric currents are confined into a small region (marked by the thick black line) close to the stellar surface (top panel); (ii) a larger, elongated region containing the currents, but still connected to the stellar surface (middle panel); and (iii) a greatly extended solution with some disconnected field lines (bottom panel). The 3D view of the latter case is shown in Fig.~\ref{fig_3d} as an illustration of a typical case. (The disconnected field lines are not shown in 3D.)

Interestingly, as we will discuss in greater length in the next section, we find that, for a fixed $s$, there is a maximum energy at some $r_{\rm c}$, beyond which the energy starts to drop. For $s=1.5$, this maximum is the model shown in the middle panel in Fig.~\ref{fig_contour} (model 2). The first disconnected configuration appears shortly after this maximum, at a critical radius of $r_{\rm c} \approx 4.69$, as becomes apparent by comparing the middle and bottom panels. Model 1 in the top panel is the lowest energy solution for $s=1.5$ and $P_{\rm c} \approx 0.57$, which, incidentally, can also be obtained through the method of Paper I, where $P_{\rm c}$ is kept fixed between iterations. The third model shown in the bottom panel has an intermediate energy, somewhat higher than model 1, but nevertheless below that of model 2.

We have carried out a thorough exploration of the parameter space. In general, for a given function $T(P)$, we have found anywhere from one, up to three solutions for the same parameters $s$ and $P_{\rm c}$. We now discuss in more detail the parameter space, and how important quantities such as energy, helicity, twist and dipole moment depend on them.

	\begin{figure*}
		\centerline{
			\includegraphics[width=1\textwidth]{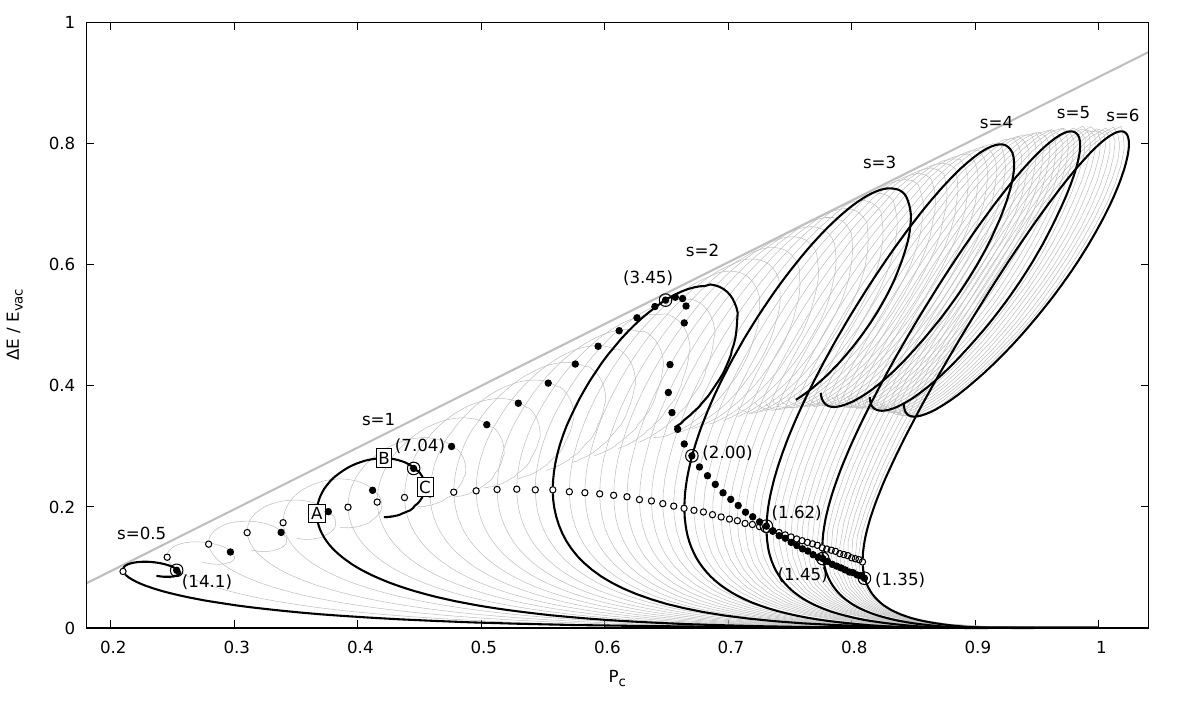}}
		\caption{Relative energy increase $\Delta E / E_{\rm vac}$ (defined in equation \ref{relative_energy}) as a function of $P_{\rm c}$. The curves are drawn for constant $s$ in the interval $0.5 \leqslant s \leqslant 6$, in increments of $0.1$. Some curves are highlighted in black for emphasis and the corresponding values of $s$ are indicated above them. For each curve, $r_{\rm c}$ varies from 1 (corresponding to the point $P_{\rm c} = 1$ at the lower right corner of the plot) up to the largest value for which reasonable convergence is achieved. The points A, B and C are indicated for the sample curve of $s=1$: A is the leftmost point along the line; B is the maximum energy; and C is the rightmost extent of the upper branch of the line, and also marks the largest value of $P_{\rm c}$ for which degenerate solutions are found for a given $s$. The leftmost points of the curves marking the separation between the high energy and low energy branches of the solutions are indicated with white circles. The points where the first disconnected field lines appear are marked with black circles for all the lines, and the corresponding values of $r_{\rm c}$ for the highlighted lines are given in parentheses.}
		\label{fig_energy}
	\end{figure*}

\subsection{Energy}
The energy stored in the magnetosphere is
	\begin{equation}
	E = \frac{1}{8\pi} \int_{R_\star}^\infty B^2 \, dV \ .
	\label{energy}
	\end{equation}
For force-free fields (including the special case of current-free fields), the energy can be expressed entirely in terms of surface integrals, as noted in Paper I. We define the relative energy with respect to the vacuum energy as
	\begin{equation}
	\Delta E = E - E_{\rm vac} \ ,
	\label{relative_energy}
	\end{equation}
where the vacuum energy for a dipole is $E_{\rm vac} = 1/3$, in the units listed in Table~\ref{table_dimensions}. In Fig.~\ref{fig_energy}, we plot the fraction $\Delta E / E_{\rm vac}$ as a function of $P_{\rm c}$, which itself is a function of $r_{\rm c}$, for various values of $s$. Note the spiral-like structure of the curves, which are plotted for constant $s$. Along each curve, $r_{\rm c}$ continuously increases starting from 1 (corresponding to the point $P_{\rm c} = 1$ near the lower right corner of the plot, and which is simply the limiting case when the toroidal field is confined to a single point on the equator) up to the largest value for which reasonable convergence is achieved. Continuing further along the curve becomes progressively difficult from a numerical perspective as $r_{\rm c}$ is increased.

	\begin{figure*}
	\centerline{\includegraphics[width=1\textwidth]{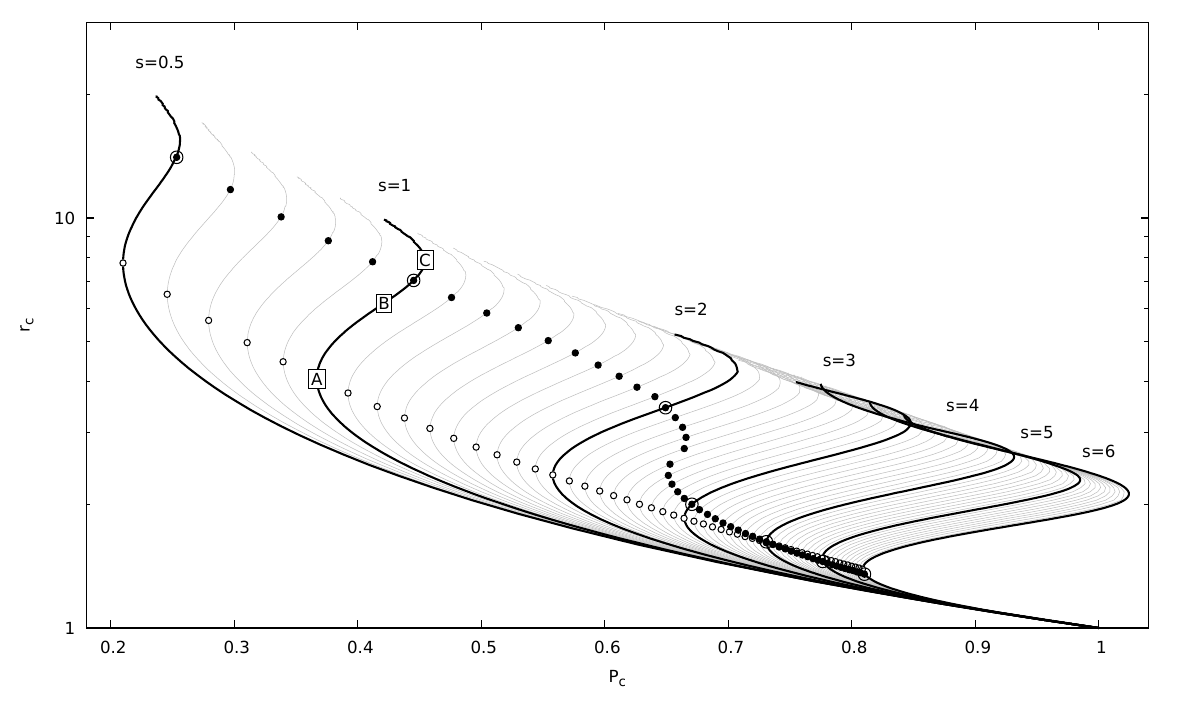}}
	\caption{$r_{\rm c}$ versus $P_{\rm c}$, for the same models as in Fig.~\ref{fig_energy}. The vertical axis is shown on a logarithmic scale in order to reveal more detail. The plot allows to convert the values of $P_{\rm c}$ to those of $r_{\rm c}$. As in Fig.~\ref{fig_energy}, the points marking the edge of the lower branch are shown as white circles, and the points where disconnected fields appear are shown as black circles.}
	\label{fig_rc}
	\end{figure*}
	
It is convenient to identify the important points along the curves. For each line of constant $s$, we define the leftmost point as \emph{point A}, the maximum energy as \emph{point B}, and the rightmost extent of the upper section of the spiral as \emph{point C}. These points are indicated for the sample curve $s=1$ in Fig.~\ref{fig_energy}. The tangents to the curve at points A and C are vertical, and at point B, it is horizontal. Point A corresponds to the lowest value of $P_{\rm c}$ for a given $s$, while point C corresponds to the largest value of $P_{\rm c}$ for which degenerate solutions are found. Thus, degenerate solutions are only present in the interval of $P_{\rm c}$ delimited by the points A and C. For larger values of $P_{\rm c}$, beyond point C, there is only one non-degenerate solution for each value of $s$.

With these definitions, we can now identify two branches for each curve: (i) the \emph{lower branch}, consisting of the lower energy solutions extending from $P_{\rm c} = 1$ (i.e.\ from $r_{\rm c} = 1$) up to point A (corresponding to some radius $r_{\rm c}$); and (ii) the \emph{upper branch}, consisting of the higher energy solutions (for larger $r_{\rm c}$). The points A for the curves of constant $s$ are shown as white circles in the figure, forming what we will refer to as \emph{line A}.

The solutions presented in Paper I correspond to the lower branch, and the highest energy configurations obtained there (for the smallest values of $P_{\rm c}$ for a given $s$) agree remarkably well with line A. (This is discussed further in \S\ref{section_map}.)

The points where disconnected domains start to appear are determined by analyzing the equatorial profile of the poloidal stream function $P$. Disconnected regions are present when $P$ has a maximum (i.e.\ its radial derivative is zero). Note that the first disconnection must be a maximum (even if there is not necessarily a minimum), because $P$ has to decrease far away from the surface, in order to smoothly connect to a vacuum field (which is of the form $P \propto r^{-l}$, where $l$ is the multipole index). The marginal disconnection takes place when the radial derivative becomes zero for the first time. The points of marginal disconnection are marked with black circles in the figure, and the corresponding values of $r_{\rm c}$ can be read from Fig.~\ref{fig_rc}, which shows the relation between $P_{\rm c}$ and $r_{\rm c}$ (allowing to translate the values of $P_{\rm c}$ into the corresponding values of $r_{\rm c}$). These values are also indicated in parentheses for the highlighted lines in Fig.~\ref{fig_energy}. Note that the formation of disconnected field lines takes place quite late along the curves for small values of $s$, while for larger values it approaches point A and may even be slightly below it. In other words, for large $s$ even the lower branch may have some disconnected fields. Either way, while such field configurations are interesting solutions of the GS equation, they may be of little practical use from a physical point of view, as they will likely result in the expulsion of a plasmoid and the sudden rearrangement of the magnetic field structure.

Finally, we note a couple of peculiarities in Fig.~\ref{fig_energy}. First, the upper branches of the curves near $s=6$ appear to exceed $P_{\rm c} = 1$ (the largest value on the stellar surface). This implies that for these models, the toroidal region is completely detached from the stellar surface, and entirely confined within a magnetospheric torus. In this case, the only lower energy solution available is the vacuum solution. Second, also note that for larger values of $s$ (in particular, for $s=5$ and $s=6$) the curve crosses itself. This point is a triply degenerate solution: the energy, as well as the parameters $s$ and $P_{\rm c}$, are the same for two solutions with different values of $r_{\rm c}$. In principle, there is no problem for the curve crossing itself, however, it does put into doubt how its subsequent continuation would be, if it could be followed further to even larger values of $r_{\rm c}$. Lastly, not far from the triply degenerate points (for the largest values of $r_{\rm c}$ for $s \gtrsim 4$), we come across the first doubly disconnected fields, that is, field configurations with two disconnected regions. In principle, it seems plausible that continuing further into ever higher values of $r_{\rm c}$ would lead to the successive appearance of multiply disconnected fields.

	\begin{figure*}
		\centerline{\includegraphics[width=1\textwidth]{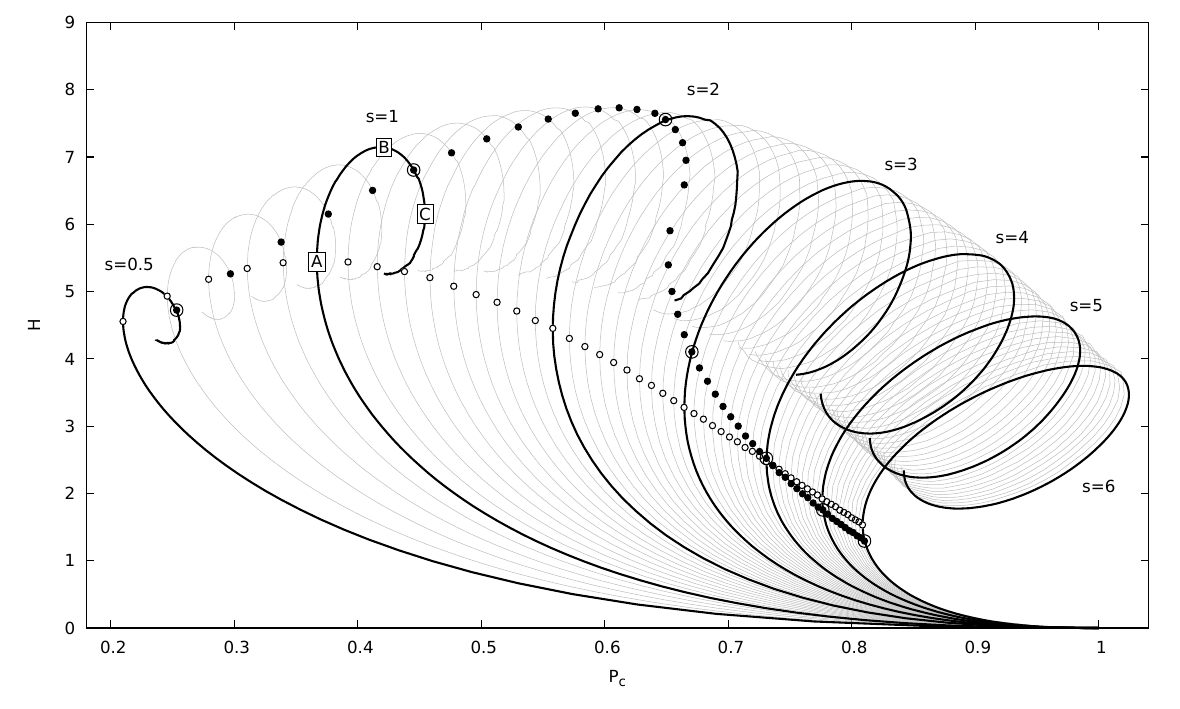}}
		\caption{Helicity $H$ versus $P_{\rm c}$. As in Fig.~\ref{fig_energy}, the curves are plotted for constant $s$, with a few of them highlighted in black lines. The leftmost points A and the points of disconnection are indicated in white and black circles, respectively. Point B corresponds to the maximum energy (for a given $s$), and is near, but not the same as, the maximum helicity.}
		\label{fig_helicity}
	\end{figure*}

\subsubsection{Implications for the energetics}
The gray line delimiting the maximum energy as a function of $P_{\rm c}$ plotted in Fig.~\ref{fig_energy} is determined through a fit by eye, and is approximately given through
	\begin{equation}
	\frac{\Delta E}{E_{\rm vac}} \approx P_{\rm c} - 0.1 \ .
	\end{equation}
In the figure, note that the largest energies attained can be up to $\sim 80\%$ larger than the ground energy level corresponding to the vacuum dipole. However, these are all in the upper branch of the curves, corresponding to the degenerate solutions, and for large $s$ they are all disconnected field configurations. Therefore, it is unlikely that these field configurations could ever be realized in nature (without some external stabilizing force). For lower values of $s$ the disconnection of field lines takes place on the upper branch (beyond point A). The largest energy attained by the (marginally) connected field configurations (indicated with black circles in the figure) is for intermediate values around $s \sim 2$ and is about $\sim 60\%$ larger than the vacuum energy. However, even these solutions may be unstable, as lower energy configurations exist for the same parameters, and it is conceivable that any perturbation would take us to the lower branch. Therefore, considering only the configurations on the lower branch (up to point A), the energy increase never exceeds the $\sim 25\%$ threshold, consistent with the results of Paper I.

Placing our results in the magnetar context, where the internal field evolution is somehow resulting in a slow, continuous injection of energy and helicity into the magnetosphere, we can argue that the maximum (magnetospheric) energy available to power a flare/outburst event is of the order of $\sim 25\%$ of the total magnetic energy of the corresponding dipole solution. Thus, in the most favorable case, this is
	\begin{equation}
	(\Delta E)_{\rm max} \approx 3.6 \times 10^{44}
	\left(\frac{B_{\rm pole}}{10^{14}\,{\rm G}}\right)^2
	\left(\frac{R_\star}{12\,{\rm km}}\right)^3 \, {\rm erg},
	\end{equation}
where $B_{\rm pole}$ is the magnetic field amplitude at the pole (i.e.\ $B_{\rm pole} = 2 B_{\rm o}$, in the units of Table~\ref{table_dimensions}), which is consistent with the observed energetics of magnetar-like events. In principle, assuming the absolute maximum value of $\sim 80\%$ could potentially triplicate this number, however, it seems unlikely that such configurations would be realized under normal circumstances.

\subsection{Helicity, twist and dipole moment}
We next consider the parametric dependence of other quantities of interest, namely the helicity, twist and dipole moment.\footnote{We refer the interested reader to Paper I, where more detailed definitions and discussions of these quantities can be found.} We define magnetic helicity as
	\begin{equation}
	H = \int \m{A}\cdot\m{B} \, dV = 2 \int A_\phi B_\phi \, dV \ .
	\label{helicity}
	\end{equation}
Here, $\m{A}$ is the vector potential, and as discussed in Paper I, the last equality is valid for a specific gauge, where a surface integration drops out.
	
On the other hand, twist is defined as the azimuthal displacement (in radians) between the footprints of a magnetospheric field line on the stellar surface. It is given through the integral
	\begin{equation}
	\varphi = \int_{0}^{\ell}
	\frac{B_\phi d\ell}{(B_r^2 + B_\theta^2)^{1/2} r\sin\theta} \ ,
	\label{twist}
	\end{equation}
where, $d\ell$ is the field line element in the ($r$,$\theta$) plane, and $\ell$ is the total length of the field line in this plane.

Finally, as in Paper I, we define the dipole strength normalized to the surface,
	\begin{equation}
	a_1 = \frac{r A_1(r)}{R_\star} \ ,
	\label{multipole_content}
	\end{equation}
where $A_1(r)$ is the dipole component obtained through the multipole expansion of the poloidal function $P(r,\theta)$ at some external radius $r > r_{\rm c}$, beyond the toroidal region containing the currents.

	\begin{figure*}
		\centerline{\includegraphics[width=1\textwidth]{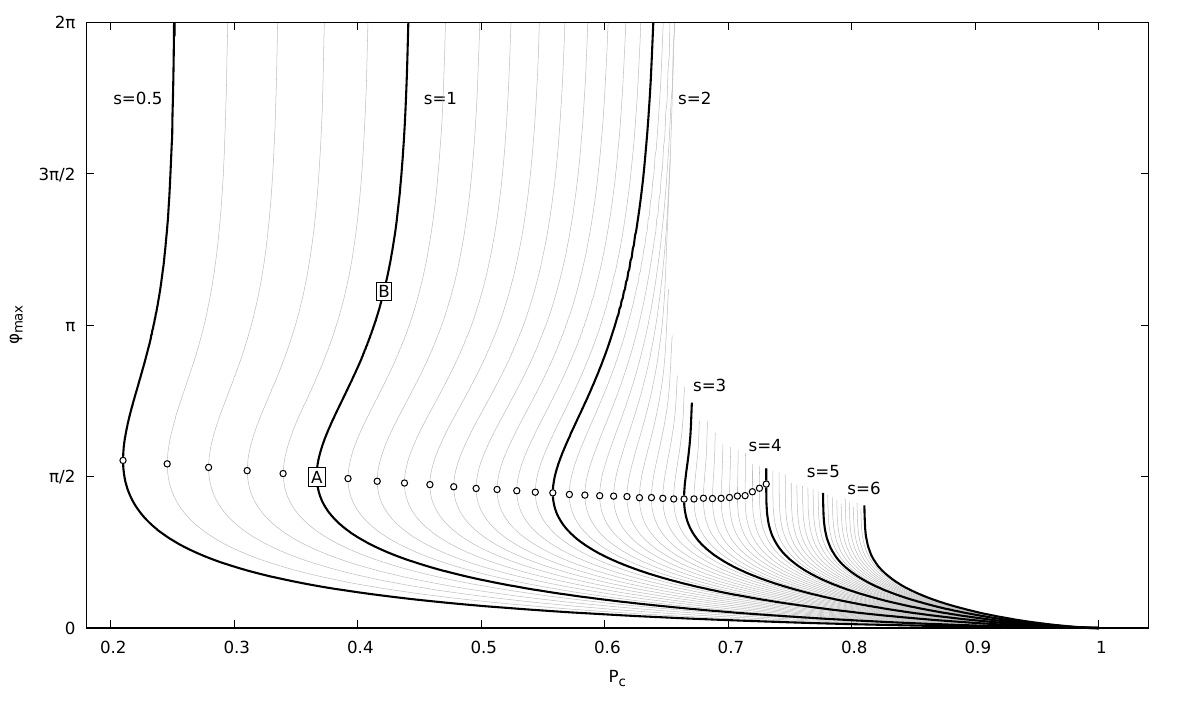}}
		\caption{Maximum twist $\varphi_{\rm max}$ versus $P_{\rm c}$. We do not calculate the twist for disconnected field lines (near the neutral point). Therefore, the lines are cut off at the points of disconnection (shown as black circles in preceding figures). Moreover, when an X-point is present for a disconnected field configuration, the twist diverges. As a consequence, the figure is capped at $2\pi$. }
		\label{fig_twist}
	\end{figure*}

\subsubsection{Helicity}
The analogous plot to Fig.~\ref{fig_energy}, but for the helicity $H$, is shown in Fig.~\ref{fig_helicity}. In this case, the points corresponding to the maximum energy (labeled as point B for $s=1$) do not necessarily correspond to the maxima of helicity, which are slightly displaced. Helicity seems to have an absolute maximum of $\sim 8$ (in the units listed in Table~\ref{table_dimensions}; see footnote\footnote{Helicity is given in units of (cf.\ Table~\ref{table_dimensions})
	\begin{equation*}
	H_{\rm o} = B_{\rm o}^2 R_\star^4 \approx 5.2 \times 10^{51}
	\left(\frac{B_{\rm pole}}{10^{14}\,{\rm G}}\right)^2
	\left(\frac{R_\star}{12\,{\rm km}}\right)^4 \, {\rm G^2\,cm^4}.
	\end{equation*}
The combination G\,cm$^2$ is equivalent to Maxwell (Mx) --- the cgs unit for magnetic flux.\label{footnote_H}}) near $P_{\rm c} \sim 0.6$. Also note that the spirals no longer cross themselves, unlike for the energy. It is not clear if this still would be the case if the curves could be continued further into larger values of $r_{\rm c}$. As in Fig.~\ref{fig_energy}, the points A are shown in white circles and the points of disconnection are shown in black circles. The solutions presented in Paper I correspond to the region below line A, which has a maximum of $\sim 5$, consistent with the conclusions of Paper I.

\subsubsection{Twist}
For each field configuration, we calculate the twist of the field lines with footprints on the stellar surface. The maximum twist --- the largest value of the integral given by equation (\ref{twist}) --- is shown in Fig.~\ref{fig_twist} as a function of $P_{\rm c}$. In Paper I, we did not find solutions beyond a maximum value of $\sim 1.5$\,rad, for a wide range of parameters ($s$ and $P_{\rm c}$). This is consistent with the white circles corresponding to the points A and marking the maximum extent of the lower branch solutions in Fig.~\ref{fig_twist}. Note how these points form a nearly horizontal line except for the largest values of $s$, effectively confirming our earlier conclusion. Thus, we can use the maximum twist as a rough indication in order to determine if a magnetospheric model is in the lower branch (for $\lesssim 1.5$) or in the upper branch (for larger values).

For the degenerate solutions of the upper branch, the twist can be significantly larger. In this case, the maximum twist is harder to calculate systematically, both because finer grids are needed in order to resolve the field lines, and because of the appearance of mathematical divergences due to the definition of the twist (through equation \ref{twist}) in the case of disconnected field lines. The latter point requires careful consideration: the definition of twist is not a straightforward matter for models with disconnected domains. In particular, the twist will diverge near the \emph{X-point}, which is a saddle point on the surface defined by $P(r,\theta)$. At this point, the radial and angular derivatives $\partial_r P$ and $\partial_\theta P$ both go to zero (while the second derivatives will have opposite signs), implying that the poloidal magnetic field components $B_r$ and $B_\theta$ simultaneously vanish, while the toroidal field $B_\phi$ and the length $\ell$ of the projection of the field line on the $(r,\theta)$ plane remain finite. From equation (\ref{twist}) for the definition of the twist, it then follows that the integrand diverges at this point and the twist goes (continuously) to infinity. A similar situation arises for the \emph{neutral point} --- the local maximum of $P$ (and the central point for the disconnected field lines), where again both partial derivatives of $P$ go to zero (while the second derivatives are both negative). However, in this case $\ell \to 0$ as well (as the projections of the field lines tend to a dot), so there may not necessarily be a divergence. We do not calculate the twist for completely disconnected field lines (near the neutral point).

	\begin{figure}
		\centerline{\includegraphics[width=0.5\textwidth]{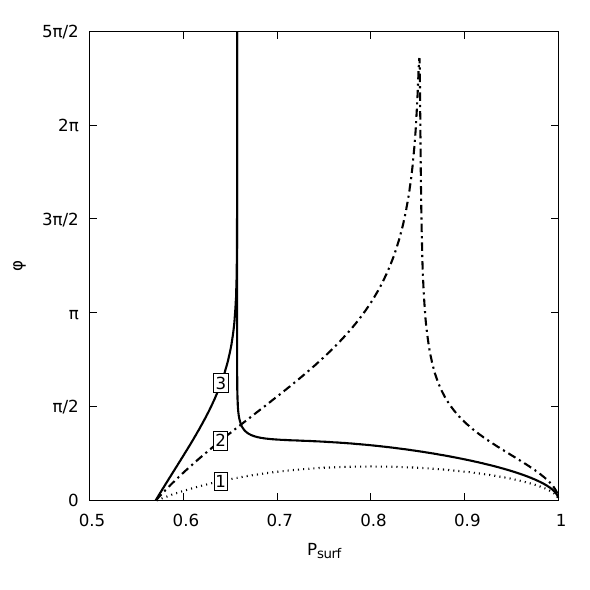}}
		\caption{Twist profiles of the three models shown in Fig.~\ref{fig_contour}. The twist $\varphi$ is shown as a function of the poloidal function at the surface $P_{\rm surf}$. When the field configuration is highly outstretched and near disconnection, as in model 2, the twist increases sharply. When eventually an X-point forms and some field lines become disconnected, the twist diverges, as in model 3.}
		\label{fig_twist_profile}
	\end{figure}

The divergence at the X-point can be understood as follows: As the poloidal field strength $B_{\rm pol}$ decreases to zero along the field line while approaching the X-point, the field line becomes more and more inclined and circular (in 3D), until, finally, at the X-point itself, $B_{\rm pol}$ vanishes and the field line turns into a circle on the equatorial plane (for a dipole field), never returning back to the surface. This case represents a purely toroidal magnetic field line, and the twist defined through equation (\ref{twist}) is no longer a particularly useful concept. Thus, when an X-point forms, there will inevitably be field lines of infinite twist, corresponding to circles with zero poloidal field strength, surrounded by nearby field lines where the  twist continuously approaches infinity. These divergences are not integrable, in the sense that they cannot be avoided by changing variables in the integration.

In Fig.~\ref{fig_twist}, for small values of $s$, the X-point forms together with a neutral point (for a sufficiently large $r_{\rm c}$) and the twist is found to diverge. For marginally disconnected fields, the twist would be large but finite, and would require a high resolution to be calculated accurately. Consequently, we choose to cut off the upper part of Fig.~\ref{fig_twist} at $2\pi$, as higher values will have larger numerical uncertainties. On the other hand, for large values of $s$, a neutral point forms initially without the presence of an X-point, and the twist of the field lines connected to the surface is still finite, while for the disconnected field lines it is not calculated. As $r_{\rm c}$ continues to increase, an X-point will eventually form as well, at which point the twist would once again diverge. In the figure, we only show the twist up to the point where the first disconnected field lines appear (corresponding to the black circles in the previous figures).

As an example, the twist profiles for the models of Fig.~\ref{fig_contour} are shown in Fig.~\ref{fig_twist_profile}. The twist $\varphi$ is shown as a function of the poloidal function at the surface, $P_{\rm surf} = \sin^2\theta$ (for dipolar boundary conditions). The twist goes to zero at $P_{\rm c}$ ($\approx 0.57$) and at 1, and has a maximum somewhere in that interval. The twist for model 1 is a smooth curve with a maximum of $\varphi_{\rm max} \approx 0.57$ (at $P_{\rm surf} \approx 0.80$). For model 2, the field lines are strongly stretched outwards, and although not yet disconnected, the X-point is about to be formed. As a consequence, the twist rises sharply to a (still finite) maximum value of $\varphi_{\rm max} \approx 7.4$ in the vicinity of $P_{\rm surf} \approx 0.85$. In model 3, an X-point has already formed and a region of disconnected field lines has appeared. Thus, the twist now diverges at the sharp cusp at $P_{\rm surf} \approx 0.66$.

	\begin{figure*}
		\centerline{\includegraphics[width=1\textwidth]{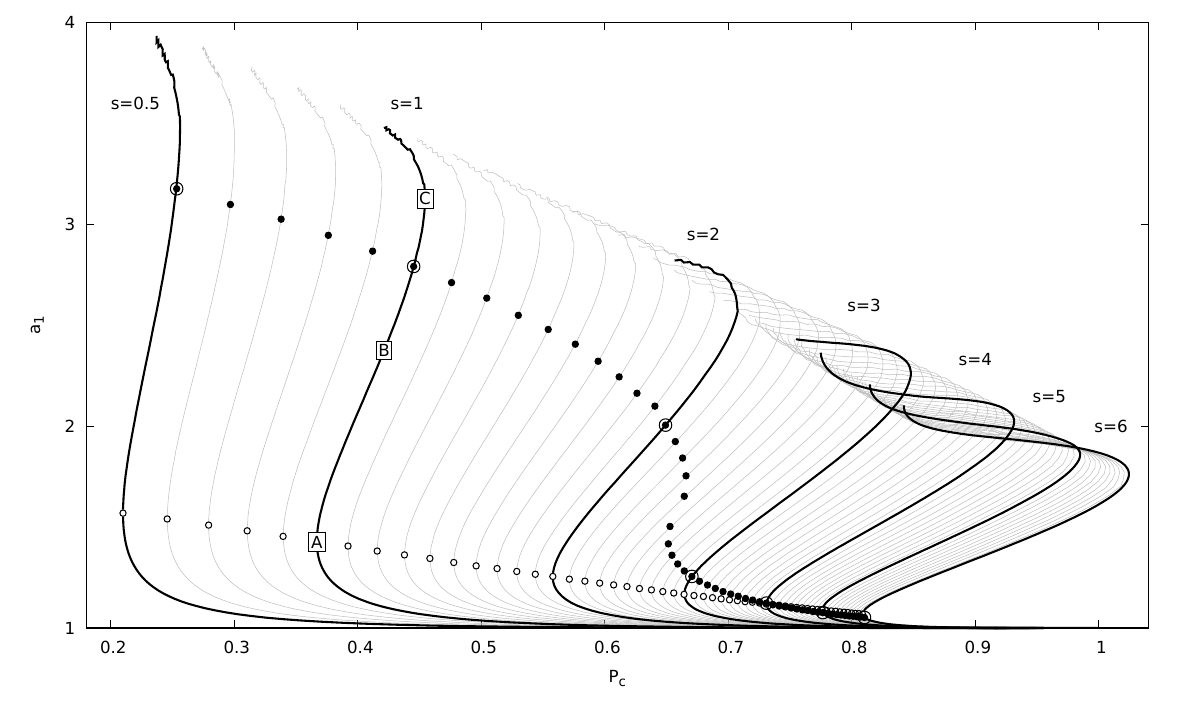}}
		\caption{Dipole strength at the surface $a_1$ versus $P_{\rm c}$.}
		\label{fig_a1}
	\end{figure*}

	\begin{figure}
		\centerline{\includegraphics[width=0.5\textwidth]{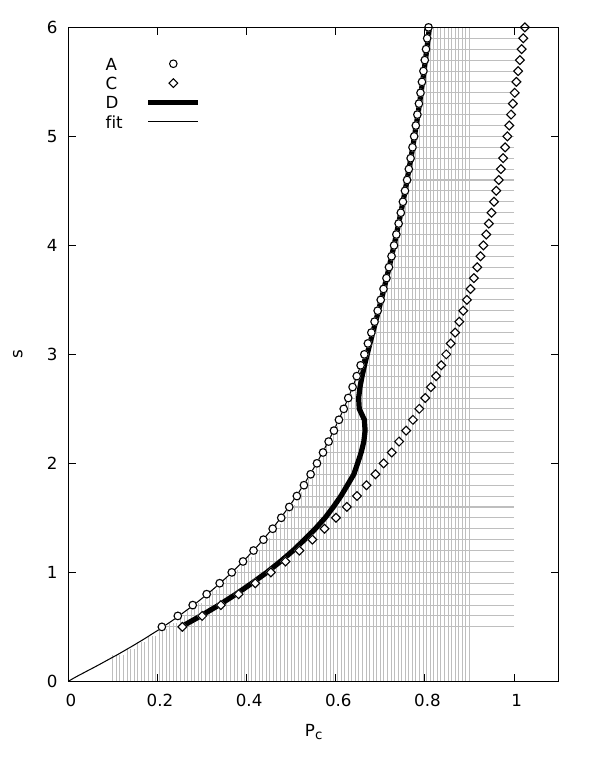}}
		\caption{Representation of the solutions in the parameter space of $s$ and $P_{\rm c}$. The vertical gray lines are the solutions found using the method of Paper I by fixing $P_{\rm c}$, while the horizontal gray lines are the solutions found with the method of this paper by fixing $r_{\rm c}$. The white space in the upper left portion is where no solutions can be found through either method. The edge of the solution space is marked by the white circles representing the points A (the smallest value of $P_{\rm c}$ for a fixed $s$). The points C representing the rightmost extent of the degenerate region are shown with white diamonds. The thick black line is for the disconnection points (shown as black circles in the previous figures, and labeled as line D here). Also shown is a fit to line A (equation \ref{fit}).}
		\label{fig_map}
	\end{figure}

\subsubsection{Dipole moment}
The dipole moment at large distances (beyond the largest extent of the currents), normalized to the value at the stellar surface (as defined through equation \ref{multipole_content}), is shown in Fig.~\ref{fig_a1}. In Paper I, we found that it can be up to $\sim 40\%$ larger than the vacuum dipole solution (which is unity in the dimensionless units used here). This is consistent with the points A indicated by the white circles. On the other hand, for the higher branch of solutions, the dipole moment can be significantly larger --- up to $4$ times larger than the vacuum case. Either way, the presence of magnetospheric currents amplifies the dipole moment, implying that measurements at large distances (for example, through the spin-down torque) overestimate the actual surface dipole moment by a factor $a_1$.

\subsection{Map of the parameter space}\label{section_map}
In Fig.~\ref{fig_map}, we show a plot of the solutions in the parameter space for $s$ and $P_{\rm c}$. The vertically shaded region corresponds to the solution space reported in Paper I, where the value of $P_{\rm c}$ was held fixed throughout the iterations. This space was explored systematically in the interval $0.1 \leqslant P_{\rm c} \leqslant 0.9$ and for values of $s$ ranging from zero up to the largest value for which solutions could be found. (Thus, this figure is analogous to Fig.~6 of Paper I.) On the other hand, the horizontally shaded region represents the solution space explored in this paper, while forcing $r_{\rm c}$ to be constant and allowing $P_{\rm c}$ to vary between iterations. In this case, the solutions have been explored in the interval $0.5 \leqslant s \leqslant 6$ and for $r_{\rm c} = 1$ (corresponding to $P_{\rm c} = 1$) up to the largest value of $r_{\rm c}$ for which convergence could be achieved.

For each horizontal line of constant $s$, the leftmost point is marked by an empty circle and corresponds to the point A, i.e.\ the lowest value of $P_{\rm c}$ for which a solution can be found. Alternatively, it also corresponds to the largest value of $s$ for a given $P_{\rm c}$ for which a solution can be found. The points A, calculated through the method used in this paper, coincide remarkably well with the largest values of $s$ obtained for the solutions presented in Paper I (shown as the vertical lines).

The white region in the upper left half of the plot beyond the boundary formed by the points A corresponds to the parameter space where no solutions through either method are found. As noted in Paper I, the edge of the parameter space, as well as contours of energy, helicity and twist, are very well approximated by a function of the form
	\begin{equation}
	s = \frac{\gamma P_{\rm c}^m}{(1 - P_{\rm c})^n} \ ,
	\label{fit}
	\end{equation}
where $\gamma$, $m$ and $n$ are three parameters to be determined by fitting. By taking the logarithm of the equation, it can be linearized. The most straightforward way is to pick three points along the boundary and solve for the unknown parameters. Taking the corresponding values of $P_{\rm c}$ for $s=1$, $2$ and $3$, we find the values $\gamma \approx 1.70$, $m \approx 0.925$ and $n \approx 0.863$ for the parameters. This function is plotted as a thin black line, and indeed, it is an outstandingly good approximation to the white circles.

The white diamonds represent the points C, which, together with the points A, delimit the region where multiple solutions can be found. To the right of line C, there is always only one solution, which is also the lowest energy solution. Note that towards the right upper corner, line C slightly goes over the value $P_{\rm c} = 1$. In these cases, the toroidal region is completely detached from the surface, and there are always only degenerate solutions, since, for the lowest energy solutions, $P_{\rm c}$ cannot exceed 1.

The thick black line shows where the first disconnected regions appear. (Shown as the black circles in the previous figures.) For small $s$, this line starts off near line C, and as $s$ is increased, it displaces to the left edge of the parameter space, eventually approaching line A.

\section{Conclusions}\label{section_conclusions}
In this work, we have extended the force-free magnetosphere solutions for the GS equation presented in Paper I (where we fixed the parameter $P_{\rm c}$ between iterations) by modifying the iteration scheme (where we now fix $r_{\rm c}$ and allow $P_{\rm c}$ to vary). As conjectured in Paper I, we show that the solutions of the GS equation are degenerate and there are multiple solutions for the same sets of parameters $s$ and $P_{\rm c}$. We are able to reproduce the solutions found previously in Paper I, which we now confirm as the lower energy (branch) solutions. In addition, we find a new branch of solutions corresponding to higher energies, some of which present disconnected domains similar to the results of \cite{2014MNRAS.445.2777F}, \cite{2015MNRAS.447.2821P} and \cite{2017MNRAS.468.2011K}.

We find that as $r_{\rm c}$ is gradually increased (starting from the stellar surface), while maintaining $s$ fixed, the field goes through the following set of configurations:
	\begin{enumerate}
	\item Initially, we are in the lower energy branch of the solutions, and the field configuration is (relatively) close to the vacuum solution. As $r_{\rm c}$ increases, the toroidal region becomes more and more inflated (as in model 1 of Fig.~\ref{fig_contour}).
	\item Beyond point A (corresponding to the white circles in Figs.~\ref{fig_energy}-\ref{fig_twist}, \ref{fig_a1} and \ref{fig_map}), degenerate solutions start to appear. Initially they are still connected to the interior, however now they present severe distortions from the vacuum case, as the field lines become elongated near the equator (as for model 2 in Fig.~\ref{fig_contour}).
	\item As $r_{\rm c}$ is increased further, the first disconnected field lines appear (beyond the marginally connected configurations indicated by black circles in Figs.~\ref{fig_energy}-\ref{fig_helicity}, \ref{fig_a1} and \ref{fig_map}). The field configurations now contain neutral points and X-points (as in model 3 of Fig.~\ref{fig_contour}; also depicted in 3D in Fig.~\ref{fig_3d}).
	\item For $s \gtrsim 4$, a second disconnected region starts to appear for the largest values of $r_{\rm c}$, implying that further solutions (with multiple disconnected regions) could exist for even larger $r_{\rm c}$. However, numerical convergence becomes progressively difficult to achieve as $r_{\rm c}$ is increased, and significantly enhancing the resolution is impractical.
	\item For sufficiently large values of $s$, the toroidal region can be completely detached from the stellar surface, as implied by $P_{\rm c} > 1$. For these extreme cases, there are no corresponding lower energy solutions, except the vacuum (current-free) solution.
	\end{enumerate}

We note that disconnected configurations are likely to be prone to severe instabilities through the ejection of a plasmoid and the sudden rearrangement of the field structure. Therefore, while these are interesting solutions of the GS equation, it is not clear how they could be naturally formed and sustained under normal circumstances in neutron stars, without some external forces. Moreover, for fixed parameters, it is conceivable that perturbations of higher energy solutions would bring the system to the lower energy configurations. Therefore, we argue that the lower branch represents the likely (more) stable solutions, while the upper branch consists of the more difficult to realize and likely unstable solutions. Disconnected field configurations are found almost entirely on the upper branch, except a narrow range for large values of $s$. Proving stability more generally requires a careful analysis of whether or not a transition to a lower energy state is at all possible, as well as an estimation of the associated timescale for such a transition, and is beyond the scope of this work.

Having identified the lower energy branch of solutions of the GS equation (roughly defined by a maximum twist of $\lesssim 1.5$\,rad, as implied by Fig.~\ref{fig_twist}), we have also determined the region of the parameter space (spanned by $s$ and $P_{\rm c}$) where solutions are possible. Interestingly, the new solutions found here do not modify the allowed parameter space reported in Paper I, described very well through a fit of the form given by equation (\ref{fit}), as depicted in Fig.~\ref{fig_map}. We argue that this border (which also corresponds to the separation of the lower and higher energy branches) is the point at which magnetospheric instabilities could be expected to produce a flare or outburst, such as the ones observed in magnetars.

Considering only the lower branch, our analysis limits the maximum energy stored in the magnetosphere to $\sim 25\%$ more than the energy of the corresponding vacuum (dipole) solution, which sets an upper bound on the energetics of the flare of about a few $10^{44}-10^{46}$\,erg for typical magnetic fields of $10^{14}-10^{15}$\,G, consistent with magnetar activity. In this case, the largest helicity is of the order of $\sim 5$ (in the units listed in Table~\ref{table_dimensions}; see footnote$^{\ref{footnote_H}}$), and the dipole strength is about $\sim 40\%$ larger than the vacuum dipole.

The degenerate upper branch allows for higher theoretical limits of up to $\sim 80\%$ more energy with respect to the vacuum case (i.e.\ about three times more than the amount for the lower branch), larger helicity of the order of $\sim 8$, and a dipole strength that can now be up to four times that of the vacuum dipole. On the other hand, the twist can be much bigger, and, in fact, it would diverge when an X-point forms in a disconnected field configuration.

This work gives support to the interpretation by \cite{2017MNRAS.472.3914A} that the criterion to determine when a flare is produced during the magneto-thermal evolution of magnetars can be determined as the point in which no solutions of the GS equation can be found. This is indeed interesting, because it allows to perform those evolutions without worrying about the complex dynamics of the magnetosphere, which instead can be substituted by the much simpler GS equation.

\section*{Acknowledgements}
This work is supported in part by the Spanish MINECO/FEDER grants AYA2015-66899-C2-1-P, AYA2015-66899-C2-2-P, the grant of Generalitat Valenciana PROMETEOII-2014-069, and by the New Compstar COST action MP1304.

\bibliographystyle{mnras}
\bibliography{references}

\label{lastpage}
\end{document}